
\magnification=1200
\documentstyle{amsppt}


\language=0


\def\SLZ{\operatorname{SL}(2,\Z)}
\def\CSLZ{{\operatorname{DSL}(2,\Z)}}
\def\MG#1{\operatorname{SL}(2,\Z/{#1}\Z)}
\def\GL#1{\operatorname{GL}(#1,\C)}
\def\GLK#1{\operatorname{GL}(#1,K)}
\def\Z{\Bbb Z}
\def\Q{\Bbb Q}
\def\C{\Bbb C}
\def\N{\Bbb N}
\def\B{\Bbb B}
\def\PH{\frak H}
\def\Im{\operatorname{Im}}
\def\Re{\operatorname{Re}}
\def\rhot{\widetilde\rho}
\def\LO{{\Cal O}}
\def\SL{\Gamma}
\def\CSL{D\Gamma}
\def\tr{\operatorname{tr}}
\def\e{\operatorname{e}}
\def\Hom{\operatorname{Hom}}
\def\End{\operatorname{End}}
\def\w{{\Cal W}}
\def\id{\operatorname{id}}

\def\hopp{\bigskip}



\def\bpz{1}
\def\witten{2}
\def\cardy{3}
\def\bouwschou{4}
\def\nrt{5}
\def\roesgen{6}
\def\UNS{7}


\def\frenkel{8}
\def\duke{9}
\def\wer{10}
\def\feher{11}
\def\zhu{12}
\def\wang{13}
\def\eho{14}
\def\andmoore{15}

\def\roc{16}
\def\capelli{17}
\def\kac{18}
\def\kacB{19}
\def\beyond{20}
\def\ehf{21}


\def\ehd{22}
\def\skoruppa{23}
\def\wohl{24}
\def\Shi{25}
\def\NobsW{26}
\def\Dorn{27}


\def\gp{28}

\font\HUGE=cmbx12 scaled \magstep4
\font\Huge=cmbx10 scaled \magstep4
\font\Large=cmr12 scaled \magstep3

\font\large=cmr17 scaled \magstep0

\document
\nopagenumbers
\pageno = 0
\centerline{\HUGE Universit\"at Bonn}
\vskip 10pt
\centerline{\Huge Physikalisches Institut}
\vskip 2.5cm
\centerline{\Large  Modular Invariance and Uniqueness}
\vskip 6pt
\centerline{\Large of Conformal Characters}
\vskip 1.6cm
\centerline{\large Wolfgang Eholzer${}^{1}$ and
                   Nils-Peter Skoruppa${}^{2}$}
\vskip 1.5cm
\vskip 1.5cm
\centerline{\bf Abstract}
\vskip 15pt
\noindent
We show that the conformal characters
of various rational models of $\w$-algebras can
be already uniquely determined if one
merely knows the central charge and the conformal
dimensions. As a side result we develop several tools
for studying representations of
$\SLZ$ on spaces of modular functions. These methods,
applied here only to certain rational conformal field
theories, may be useful for the analysis of many others.
\vfill
\settabs \+&  \hskip 110mm & \phantom{XXXXXXXXXXX} & \cr
\+ & Post address:                       & BONN-TH-94-16   & \cr
\+ & ${}^1$Nu{\ss}allee 12               & MPI-94-67       & \cr
\+ & 53115  Bonn, Germany                & hep-th/9407074  & \cr
\+ & ${}^2$ UFR de Math\'ematiques et
     Informatiques                       & July 1994       & \cr
\+ & 351 rue de la Lib\'eration,         &                 & \cr
\+ & 33405 Talence, France               &                 & \cr\eject
\eject

\topmatter
        \address
Max-Planck-Institut f\"ur Mathematik Bonn,
Gottfried-Claren-Stra{\ss}e 26,
53225 Bonn, Germany
        \endaddress
\email eholzer\@mpim-bonn.mpg.de \endemail
        \address
Physikalisches Institut der Universit\"at Bonn,
Nu{\ss}allee 12,
53115 Bonn, Germany
        \endaddress
\email eholzer\@avzw01.physik.uni-bonn.de \endemail
        \address
Universit\'e Bordeaux I,
UFR de Math\'ematiques et Informatiques,
351 rue de la Lib\'eration,
33405 Talence, France
        \endaddress
\email skoruppa\@ceremab.u-bordeaux.fr \endemail

	\toc
\head
1. Introduction
\endhead
\head
2. Vertex operator algebras, $\w$-algebras
   and rational models
\endhead
\head
3. Central charges and conformal dimensions
   of certain rational models
\endhead
\head
4. Uniqueness of conformal characters
   of certain rational models
\endhead
\subhead
4.1 Statement of the main theorem
\endsubhead
\subhead
4.2 A dimension formula for vector valued modular forms
\endsubhead
\subhead
4.3 Three basic lemmas on representations of $\SLZ$
\endsubhead
\subhead
4.4 Proof of the main theorem
\endsubhead
	\endtoc

\endtopmatter

\rightheadtext{Modular invariance and uniqueness of
               conformal characters}
\leftheadtext{Wolfgang Eholzer and Nils-Peter Skoruppa}


\head
1. Introduction
\endhead

In the last years two-dimensional conformal field theories
played a profound role in theoretical physics as well as in
mathematics. Starting with the work of A.A. Belavin, A.M. Polyakov
and  A.B. Zamolodchikov \cite{\bpz} in 1984, many
new results connecting statistical mechanics and string theory
with the theory of topological invariants of 3-manifolds or with
number theory were found \cite{\witten}\cite{\cardy}.
In mathematical physics the classification of rational conformal field
theories (RCFT) became one of the important outstanding problems.

Since one hopes that it is possible to consider all RCFTs as
rational models of $\w$-algebras, special vertex operator algebras
generalizing in a certain sense Kac-Moody algebras,
different methods for the investigation of these algebras  and their
representations have been developed (for a review see e.g.\ \cite{\bouwschou}).

An important tool in the study of rational models of $\w$-algebras
are the associated conformal characters.
These conformal characters $\chi_h$ form a finite set of modular functions
satisfying a transformation law
$$\chi_h(A\tau)=\sum_{h'} \rho(A)_{h,h'}\chi_{h'}(\tau).$$
Here $A$ runs through the full modular group $\SL=\SLZ$ or through a
certain subgroup $G(2)$ (accordingly as the underlying $\w$-algebra is
bosonic or fermionic), and $\rho$ is a matrix representation of
$\Gamma$ or $G(2)$, which depends on the rational model
under consideration.

It has already been noticed that conformal characters are very
distinguished modular functions: First of all,
similar to the $j$-function, their Fourier coefficients are nonnegative
integers and they have no poles in the upper half plane. They
sometimes admit interesting sum formulas: These formulas, which
allow an interpretation as
generating functions of the spectrum of certain quasi-particles, can
be used to
deduce dilogarithm-identities (see e.~g.\ \cite{\nrt,\roesgen}). In some
cases the conformal characters
have
simple product expansions. If one has both, sum and product expansions, the
resulting identities are what is known in combinatorics as Roger-Ramanujan or,
more general, as  Andrews-Gordon identities.

In this paper we add one more piece to this theme.
We show, for certain rational models, that the central charge
and the finite set of conformal dimensions uniquely determine
its conformal characters. More precisely, we shall state a few  general
and simple axioms which are satisfied by the conformal characters of all
known rational models of $\w$-algebras. These axioms state essentially
not more than the $\SLZ$-invariance of the space of functions spanned by
 the conformal characters, the rationality of their Fourier coefficients
and an upper bound for the order of their poles. The only data of
the underlying rational model occurring in these axioms are the central
charge and the conformal dimensions, which give the upper bound for the
pole orders and a certain restriction on the $\SLZ$-invariance.
We then prove that, for various sets of central charges and conformal
dimensions, there is at most one set of modular  functions which satisfies
these axioms (cf\. the Main Theorem in~\S~4).

This result has several implications. First, it shows that the simple
constraints imposed on modular functions by the indicated axioms are
surprisingly restrictive. Apart from giving an aesthetical satisfaction
this observation gives further evidence that conformal characters are modular
functions of a rather special nature, which may deserve further studies,
even independently of the theory of $\w$-algebras.

Secondly, it implies that, in the case of the rational models considered in
this article, the conformal characters do a priori not give more
information about the underlying rational model than
the central charge and the conformal dimensions.
This is in perfect accordance with the more general belief that
these data already determine completely the rational models of
$\w$-algebras which do not contain currents (currents are nonzero elements
of dimension 1; see \S 2). In general
one expects that a unique characterization of rational models can
be obtained if one takes into account certain
additional quantum numbers  which can be defined in terms of the Lie algebra
spanned by the zero modes of the currents.

Thirdly, our main result has a useful practical consequence for the
computation of conformal characters.
Apart from several well-understood rational models where one has simple closed
formulas for the conformal characters, it is in general difficult to compute
them directly. Any attempt to obtain the first few Fourier coefficients
by the so-called direct calculations in the $\w$-algebra, the so far only known
method in the case where no closed formulas are available, requires
considerable
computer power.  Our result indicates a way to avoid the direct calculations:
Once the central charge and conformal dimensions are determined the computation
of the conformal characters can be viewed as a problem which belongs solely to
the theory of modular forms, i.e\. a problem whose solution affords no further
data of the rational model in question.
We shall show elsewhere how one can indeed solve this problem in many cases
using theta series, and, in particular,
how one obtains in this way explicit closed formulas for
the conformal characters of certain nontrivial models
which could  not be computed using known methods
\cite{\UNS}.

In this paper we restrict our attention to rational models of $\w$-algebras
where the associated representation $\rho$
turns out to be irreducible. This restriction is mainly
of technical nature: It simplifies the identification of $\rho$.
However, we believe that the Main Theorem holds true
in more generality, i.e\. that it can be extended to rational
models  with composite $\rho$, possibly with a
slightly larger set of axioms.

We have organized our article as follows:
In \S 2 we give (axiomatic) definitions of the basic
notions concerning $\w$-algebras since there seems to be no
satisfactory reference for this.
In \S 3 we give a short overview of those rational models
for which we prove our Main Theorem.
We do not feel competent to judge the literature cited in this section to
be acceptable by a physicist as well as a mathematician in view of its
mathematical cleanness, and there might be a dispute whether the existence of
various rational models mentioned in \S 3 is rigorously proved or not. Our
policy here is that we simply cite what is asserted in the literature.
What is actually needed from this (short) section are solely the Tables 1 and
2.
In \S 4 we state and prove our main result. The sections
\S 4.2 and \S 4.3, where we develop the necessary tools needed for the proof
of the Main Theorem,
may be of independent interest for those studying representations $\rho$
arising from conformal characters.

\subhead
Notation
\endsubhead
We use $\PH$ for the complex upper half plane, $\tau$ as a
variable in $\PH$, $q= e^{2\pi i \tau}$,
$$T=\left(\smallmatrix 1&1\\0&1\endsmallmatrix\right),\qquad
S=\left(\smallmatrix 0&-1\\1&0\endsmallmatrix\right),$$
$\SL$ for the group $\SLZ$, and
$$\Gamma(n) = \{ A \in \SLZ \ \vert \ A \equiv \id \pmod n \}  $$
for the principal  congruence subgroup  of $\SLZ$ of level $n$.
We use  $\eta$ for the Dedekind eta function
$$\eta(\tau)=\e^{\pi i\tau/12}\prod_{n\ge 1}(1-q^n).$$


\head
2. Vertex operator algebras, $\w$-algebras and rational models
\endhead

$\w$-algebras are a special kind of vertex operator algebras.
For the reader's convenience we repeat the definition of vertex
operator algebras and their representations
(see e.g.\ \cite{\frenkel,\duke}).

\definition{Definition (Vertex operator algebra)}
A vertex operator algebra is a complex $\N$-graded vector space
$$ V = \bigoplus_{n\in\N} V_n$$
with $\dim(V_n) < \infty$ for all $n\in\N $
(an element $\phi\in V_n$ is said to be of dimension $n$),
together with a linear map
$$V \to (\End V)[[z,z^{-1}]],\qquad
  \phi\mapsto Y(\phi,z) = \sum_{n\in\Z} \phi_n\ z^{-n-1}, $$
(the elements of the image are called vertex operators), and
two distinguished elements $1 \in V_0$ (called the vacuum)
and $\omega\in V_2$ (called the Virasoro element)
satisfying the following axioms:
\roster
\item
The map $\phi \mapsto Y(\phi,z)$ is injective.
\item
For all $\phi,\psi\in V$ there exists a $n_0$ such that
$\phi_n \psi = 0$ for all $n\ge n_0$.
\item
For all $\phi,\psi\in V$ and $m,n\in\Z$ one has
$$
  (\phi_m \psi)_n = \sum_{i\ge 0} (-1)^i \binom{m}{i}
   \left( \phi_{m-i} \psi_{n+i} -
          (-1)^m \psi_{m+n-i} \phi_{i} \right). $$
For $m<0$ this identity has to be read argumentwise: Note that by (2) the
sum on the right hand side becomes finite when
applied to an element of $V$.
\item
$Y(1,z) = \id_V$.
\item
Writing $Y(\omega,z) = \sum_{n\in\Z} L_{n}z^{-n-2}$ one has
$$ L_0\restriction{V_n} = n\,\id_{V_n}, $$
$$ Y(L_{-1}\phi,z) = \frac{d}{dz}Y(\phi,z),$$
$$ [L_m,L_n] = (m-n)L_{m+n} +
   \delta_{m+n,0}\, (m^3-m) \frac{c}{12} \id_V,$$
for all $n,m\in\Z$, $\phi\in V$, where $c$
is a complex constant (called the central charge or rank).
\endroster
\enddefinition

\remark{Remarks}
1. For $m\ge 0$ property (5) is equivalent to
$$ [\psi_m,\phi_n] = \sum_{i\ge 0} \binom{m}{i}
                                   (\psi_i\phi)_{m+n-i}. $$
where the left hand side denotes the ordinary commutator
of endomorphisms.

2. This commutator identity implies in particular
$[L_0,\phi_n]=(L_{-1}\phi)_{n+1}+(L_{0}\phi)_n$, hence
$[L_0,\phi_n]=(d-n-1)\phi_n$
for $\phi\in V_d$ (here we used $(L_{-1}\phi)_{n+1}=(-n-1)\phi_n$
from axiom~(4)). From this one obtains
$$\phi_nV_m\subseteq V_{m+d-n-1}.$$
\endremark

\definition{Definition (Representation of a vertex operator algebra)}
A representation of a vertex operator algebra $V$
is a
linear map
$$\rho\colon V \to (\End M)[[z,z^{-1}]],\qquad
   \phi\mapsto Y_M(\phi,z) = \sum_{n\in\Z} \rho(\phi)_n z^{-n-1}, $$
where $M$ is a
$\N$-graded complex vector space
$$ M = \bigoplus_{n\in\N} M_n$$
with $\dim(M_n) < \infty$ for all $n\in\N$,
such that the following axioms are satisfied:
\roster
\item
For all $\phi\in V_d$ and $m,n$ one has
$\rho(\phi)_n M_m \subset M_{m-n-1+d}.$
\item
For all $\phi\in V$ and $v\in M$ there exist a $n_0$ such that
$\rho(\phi)_n v = 0$
for all $n\ge n_0$.
\item
For all $\phi, \psi\in V$ and all $m,n\in\Z$ one has
$$ \rho(\phi_m \psi)_n = \sum_{i\ge 0} (-1)^i \binom{m}{i}
   \left( \rho(\phi)_{m-i} \rho(\psi)_{n+i} -
          (-1)^m \rho(\psi)_{m+n-i} \rho(\phi)_{i} \right), $$
where again this identity has to read
argumentwise.
\item
$Y_M(1,z) =  \id_{M}.$
\item
Using $Y_M(\omega,z) = \sum_{n\in \Z} \rho(L)_n z^{-n-2}$ one has
$$ Y_M(L_{-1}\phi,z) = \frac{d}{dz}Y_M(\phi,z),$$
$$ [\rho(L)_m,\rho(L)_n] = (m-n)\rho(L)_{m+n} +
   \delta_{m+n,0}\, (m^3-m) \frac{c}{12} \id_M,$$
for all $n,m\in\Z$, $\phi\in V$,
where $c$ is the central charge of $V$.
\endroster
The representation $\rho$ is called irreducible if there is
no nontrivial subspace of $M$ which is invariant under all
$\rho(\phi)_n$.
\enddefinition
In the following we shall occasionally use simply the term
$V$-module $M$ instead of representation $\rho\:V@>>>\End(M)[[z,z^{-1}]]$.
\remark{Remarks}
Note that a vertex operator algebra $V$ is a $V$-module itself
via $\phi\mapsto Y(\phi,z)$
(use remark~(2) after the definition of vertex operator algebra
for verifying axiom~(1) of a representation).
\endremark

\proclaim{Lemma}
Let $\rho\colon V@>>>\End(M)[[z,z^{-1}]]$ be an irreducible
representation of the vertex operator
algebra $V$. Then there exists a complex constant $h_m$ such that
$$ \rho(L)_0 \restriction{M_n} = (h_M+n)\,\id_{M_n}$$
for all $n\in\N$.
\endproclaim
\demo{Proof}
By axiom (1) of a vertex operator algebra representation we have that
$\rho(L)_0M_0\subseteq M_0$.
Hence, since $M_0$ is finite dimensional, there exists an eigenvector $v$ of
$\rho(L)_0$ in $M_0$. Let $h_M$ be the corresponding eigenvalue.
Since $\rho$ is irreducible the vector space $M$ is generated by the
vectors $\rho(\phi)_nv$ ($\phi\i V_d$, $d\in\N$, $n\in\Z$); for proving
this note that the subspace spanned by the latter vectors is invariant
under all $\rho(\phi)_n$ as can be deduced from axiom~(3)). For $m\in\N$
let $M_m^\prime$ be the subspace generated by all $\rho(\phi)_nv$ with
$\phi\in M_d$ and $d-n-1=m$.
By axiom (1) we have $M_m^\prime\subseteq M_m$, and since $M$ is the
sum of all
the $M_m^\prime$ we conclude $M_m^\prime=M_m$.

On the other hand, one has $[\rho(L)_0,\rho(\phi)_n]=(d-n-1)\phi_n$ for all
 $n$ and all $\phi\in V_d$ (similarly as in remark~(2) after the definition
of vertex operator algebras). From this we obtain
$\rho(L)\restriction M_n^\prime=(h_M+n)\,\id_{M_n^\prime}$. This proves the
lemma.
\qed\enddemo

The lemma suggests the following
\definition{ Definition (Character of a vertex operator algebra module)}
Let  $M$ be an irreducible module of the vertex
operator algebra $V$ (with respect to the representation $\rho$).
Then the character $\chi_M$ of $M$ is
the formal power series defined by
$$ \chi_M(q) := \tr_{M}( q^{\rho(L)_0-c/24} )
           := q^{h_M-c/24}\sum_{n\in\N} \dim(M_n) q^{n}
$$
where $c$ is the central charge of $V$ and $h_M$ the
conformal dimension of $M$.
\enddefinition

The most important class of vertex operator algebras if
given by ``rational'' vertex operator algebras:
\definition{Definition (Rationality of vertex operator algebras)}
A vertex operator algebra $V$ is called rational if
the following axioms are satisfied:
\roster
\item
$V$ has only finitely many inequivalent irreducible representations.
\item
Every finitely generated representation of $V$ is equivalent
to a direct sum of finitely many irreducible representations.
\endroster
\enddefinition

Here the notions equivalence, finitely generated and direct sum
are to be understood in the obvious sense. The importance of the rational
algebras becomes clear by the following theorem:

\proclaim{ Theorem (Zhu \cite{\zhu})}
Let $M_i$ ($i=1,\dots,n$) be a complete set irreducible
modules of the rational vertex operator algebra $V$.
Assume, furthermore, that Zhu's  finiteness condition
is satisfied, i.e.\
$$ \dim( V/(V)_{-2}V ) < \infty $$
where $(V)_{-2}V \subset V$ is defined by
$(V)_{-2}V := \{ \phi_{-2} \psi \vert \phi,\psi\in V\}.$
Then the conformal characters $\chi_{M_i}$ become holomorphic
functions on the upper complex half $\PH$ plane by setting
$q = e^{2\pi i\tau}$ with $\tau\in\PH$.
Furthermore, the space spanned by the conformal characters
$\chi_{M_i}$ ($i=1,\dots,n$) is invariant under the natural
action $(\chi(\tau),A) \mapsto \chi(A\tau)$
of the modular group $\SLZ$.
\endproclaim

We now turn to the definition of $\w$-algebras and rational
models of $\w$-algebras. As indicated above
we describe these in terms of vertex operator algebras.

\definition{Definition ($\w$-algebra)}
A vertex operator algebra $V$ is called
a (bosonic) $\w$-algebra
if it satisfies the following additional axioms:
\roster
\item
$\dim(V_0) = 1.$
\item
There exist finitely many homogeneous elements $\phi^i\in \ker(L_1)$
($i=1,\dots,n$)  which generate $V$.
\endroster
\enddefinition
Here vectors $\phi^i$ ($i=1,\dots,n$) are said to generate $V$
if the smallest subspace of $V$ which is invariant under the
action of $(\phi^i)_m$ ($i=1,\dots,n; m\in\Z$) and contains $1$
equals $V$.

A $\w$-algebra $V$ is said to be of type $\w(d_1,\dots,d_n)$
if there exists a minimal set of homogeneous generators
$\phi^i\in \ker(L_1) $ ($i=1,\dots,n$) whose dimensions equal
$d_1,\dots,d_n$. Here minimal means that no proper subset of
the set of the $\phi^i$ generates $V$.
Note that the $d_i$ occurring here may in general not be unique.

\remark{Remarks}
1. Examples of $\w$-algebras can be constructed from the Virasoro
and Kac-Moody algebras. They are of type $\w(1,\dots,1)$,
respectively $\w(2)$ for the Virasoro algebra \cite{\duke}.

2. Note the following for connecting our definition of $\w$-algebras
with the corresponding notion used in the physical literature.
The right hand side of (5) in the definition of vertex operator
algebras is, for $m<0$, what is usually called the $n$-th mode
$N(\psi,\partial^{-1-m}\phi)_n$ of the normal ordered product of the
vertex operators corresponding to $\psi$
and the $(-m-1)$-th derivative of vertex operator corresponding to $\phi$
(see e.g.\ \cite{\wer})
Moreover, the commutator formula in remark (1) after the definition of
vertex operator algebras implies
the (in the physical literature)
well-known formula for the commutator of two homogeneous
elements in $\ker(L_1)$ of a $\w$-algebra $V$
(see e.g.\ \cite{\wer,\feher}).
\endremark

\definition{Definition (Rational model)}
A rational model (or rational model of a $\w$-algebra) is a
rational $\w$-algebra $V$ which
satisfies Zhu's finiteness condition.
The {\it effective central charge} of a rational model is
defined by
$${\tilde c} = c - 24 \min h_{M_i} $$
where $M_i$ runs through a complete set of inequivalent irreducible
representations of $V$.
\enddefinition
\remark{Remarks}
1. Examples of rational models are given by
certain vertex operator algebras constructed from Kac-Moody
algebras \cite{\duke} or the Virasoro algebra \cite{\wang}
(for more details see also \S 3).

2. One can show that the effective central charge of a rational
model with a minimal generating set of $n$ vectors lies in the
range \cite{\eho}
$$ 0 \le {\tilde c}  < n.$$

3. Historically the term ``rational models'' was used in the physical
literature \cite{\bpz} for field theories in which the operator
product expansion of any two local quantum fields decomposes into
finitely many conformal families.
\endremark

The following theorem justifies the terminology ``rational models'':

\proclaim{Theorem (\cite{\andmoore}) }
Assume that the representation of the modular group acting on
the space spanned by the conformal characters of a rational
model is unitary.
Then the central charge and the conformal dimensions of the
rational model are rational numbers.
\endproclaim


\head
3. Central charges and conformal dimensions of certain rational models
\endhead

In this section we review some facts about those rational models
which are concerned by the Main Theorem in \S 4.
Note that some of the results summarized in this section are
not yet proved on a rigorous mathematical level.
However, we shall not be concerned by this since we are only
interested in the central charges and sets of conformal dimensions
provided by these models. This section serves rather as a
motivation than as a background for the considerations in the
subsequent sections.

Firstly, we review some known rational models with effective
central charge less than 1.
The simplest $\w$-algebras are those which can be constructed from
the Virasoro algebra (as already mentioned in the foregoing section).
 The rational models among these are called
the Virasoro minimal models (see e.g.\ \cite{\bpz, \roc,\wang}).
They can be parameterized by a set of two coprime integers
$p,q\ge 2$.
The rational model corresponding to such a
set $p,q$ has central charge
$$ c = c(p,q) = 1-6\frac{(p-q)^2}{pq} $$
and its conformal dimensions  are given by:
$$ h(p,q,r,s) = \frac{(rp-sq)^2-(p-q)^2}{4pq}
   \quad (1\le r<q,\ (2,r)=1, \ s=1\le s<p),
$$
where we assume $q$ to be odd.
\eject
The Virasoro minimal models are special examples of the larger
class of rational models with $\tilde c < 1$ which emerges from
the $ADE$-classification  of modular invariant partition functions
\cite{\capelli,\eho}.
Their central charges and conformal dimensions are given in Table 1:
The first column describes the type of modular invariant
partition function, the central charge is always $c= c(p,q)$
where $p$ and $q$ are the parameters of the respective row
under consideration. Moreover, $c(p,q)$ and $h(p,q,\cdot,\cdot)$
are as defined above.
Note that the listed models exist also for $p,q,m$ not
necessary prime. The primality restrictions have been added for
technical reasons only which will become clear in the next section.

\hopp
\centerline{Table 1: Data of certain $\w$-algebras related
                     to the $ADE$-classification}
\smallskip\noindent
\centerline{
\vbox{ \offinterlineskip
\def\tablespace{ height2pt&\omit&&\omit&&\omit&\cr }
\def\tablerule{ \tablespace
                \noalign{\hrule}
                \tablespace      }
\hrule
\halign{&\vrule#&
  \strut\quad\hfil#\hfil\quad\cr
\tablespace
& type
  && type of $\w$-algebra
  && $H_{c(p,q)}\qquad (I_n := \{1,\dots,n\})$ &\cr
\tablerule
& $(A_{q-1},A_{p-1})$
  &&   $ \w(2)$
  &&  $\{ h(p,q,r,s) \ \vert \ r\in I_{q-1},
                             \ s\in I_{p-1},\ (2,r)=1\}$
&   \cr \tablespace
&\omit
  && $p>q$  prime
  &&\omit
&\cr \tablerule
&$(A_{q-1},D_{m+1})$
  &&   $\w(2,\frac{(m-1)(q-2)}{2})$
  &&  $\{ h(p,q,r,s) \ \vert \ r\in I_{(q-1)/2},
                             \ s\in I_m,\ (2,s)=1 \}$
&   \cr \tablespace
&\omit
  && $p=2m$
  && \omit
&\cr \tablespace
&\omit
  && $q,m$ prime
  &&\omit
&\cr \tablerule
& $(A_{q-1},E_6)$
  &&   $\w(2,q-3)$
  &&  $\{ \min(h(p,q,r,1),h(p,q,r,\ 7)) \ \vert
          \ r\in I_{(q-1)/2} \}  \cup$
& \cr\tablespace
&\omit
  && $p=12,\ q\ge5$
  &&  $\{ \min(h(p,q,r,5),h(p,q,r,11)) \ \vert
          \ r\in I_{(q-1)/2} \}  \cup$
& \cr\tablespace
& \omit
  &&  $q$ prime
  && $ \{ h(p,q,r,4) \ \vert \ r\in I_{(q-1)/2} \}$
&   \cr  \tablerule
&$(A_{q-1},E_8)$
  &&   $\w(2,q-5)$
  &&  $\ \{ \min(h(p,q,r,1),h(p,q,r,11)) \ \vert
        \ r\in I_{(q-1)/2} \}  \cup$
& \cr\tablespace
&\omit
  && $p=30,\ q\ge7$
  && $\{ \min(h(p,q,r,7),h(p,q,r,13)) \ \vert
           \ r\in I_{(q-1)/2} \}\ $
&   \cr \tablespace
&\omit
  &&$q$ prime
  && \omit
&\cr \tablespace
}
\hrule}
}

The second list of rational models which we shall consider
are special cases of the so-called Casimir $\w$-algebras.

Starting from a Kac-Moody algebra associated to a simple Lie algebra
${\Cal K}$ one can construct a 1-parameter family $\w\Cal K$ of
$\w$-algebras, the parameter being the central charge
(see e.g.\ \cite{\kac}) (Note that this construction
is different from the one mentioned in the foregoing section).
For all but a finite number of central charges these $\w$-algebras
are of type $\w(d_1,\dots,d_n)$ where  $n$ is the rank of $\Cal K$ and
the $d_i$ ($i=1,\dots,n$) are the orders of the Casimir
operators of $\Cal K$.
The remaining ones, called truncated, are of type
$\w(d_{i_1},\dots,d_{i_r})$  where the $d_{i_k}$ form a proper
subfamily of the $d_i$ above.
Note that the $\w$-algebras constructed from the Virasoro algebra
mentioned in \S 2 are exactly the  Casimir $\w$-algebras
associated to ${\Cal A}_1$.
The rational models of Casimir $\w$-algebras
(sometimes called minimal models) have been determined,
assuming certain conjectures, in \cite{\kac}.

In Table 2 we list the central charges $c$, effective central
charge $\tilde c$ and the sets of conformal dimensions $H_c$  of
$6$ rational models  with $\tilde c > 1$.

The last four are Casimir $\w$-algebras associated to
${\Cal B}_2,{\Cal G}_2,{\Cal E}_7$ and ${\Cal B}_3$.

The first two $\w$-algebras are ``tensor products'' of
the rational $\w$-algebra with $c=-22/5$ constructed
from the Virasoro algebra and the rational $\w$-algebras
with $c= 14/5$ or $c=26/5$ constructed from the Kac-Moody
algebras associated to  ${\Cal G}_2$ or ${\Cal F}_4$, respectively.
We denote them by $\w_{{\Cal G}_2}(2,1^{14})$ and
$\w_{{\Cal F}_4}(2,1^{14})$, respectively.
Here the construction of the $\w$-algebras in question is the
one mentioned in \S 2.

\hopp
\centerline{Table 2:  Data of the six rational models }
\smallskip\noindent
\centerline{
\vbox{ \offinterlineskip
\def\tablespace{ height2pt&\omit&&\omit&&\omit&&\omit&\cr }
\def\tablerule{ \tablespace
                \noalign{\hrule}
                \tablespace      }
\hrule
\halign{&\vrule#&
  \strut\quad\hfil#\hfil\quad\cr
\tablespace
& $\w$-algebra && $c$    && $\tilde{c}$  && $H_c$ &\cr
\tablerule
& $\w_{G_2}(2,1^{14})$  &&  $-{8\over5}$  && ${16\over5}$ &&
  $ {1\over5} \{0,-1,1,2 \} $
                    &\cr \tablerule
& $\w_{F_4}(2,1^{26})$  &&  ${4\over5}$  && ${28\over5}$ &&
  $ {1\over5} \{0,-1,2,3 \} $
                    &\cr \tablerule
& $\w(2,4)$  &&  $-{444\over11}$  && ${12\over11}$ &&
  $ -{1\over11}
               \{0,9,10,12,14,15,16,17,18,19  \} $
                    &\cr \tablerule
& $\w(2,6)$  &&  $-{1420\over17}$ && ${20\over17}$ &&
  $ -{1\over17}
               \{0,27,30,37,39,46,48,49,50,$&\cr
&\omit  && \omit    && \omit             &&
           \qquad $52,53,55,57,58,59,60  \} $
                    &\cr \tablerule
& $\w(2,8)$  &&  $-{3164\over23}$ && ${28\over23}$ &&
  $-{1\over23}
               \{ 0,54,67,81,91,94,98,103,111,$&\cr
&\omit      && \omit   &&\omit           && \quad
                $112,116,118,119,120,122,124, $ &\cr
&\omit       && \omit  &&\omit           &&
          \quad $125,129,130,131,132,133 \} $
                    &\cr \tablerule
& $\w(2,4,6)$  &&  $-{13\over15}$ && ${17\over15}$ &&
  $ {1\over180} \{0, -15, -8, -3, 12, 37, 57, 60, 100,$&\cr
& \omit  && \omit    && \omit             &&
           \qquad $ 117, 120, 132, 145, 252, 285, 405 \} $
                    &\cr \tablespace}
\hrule}
}

We give some comments on these $6$ rational models.
Using \cite{\roc} and \cite{\kacB} the central charges, conformal
characters and  dimensions of the two composite rational models
can be computed. For the rational models of type
$\w(2,d)$   lists of the associated conformal dimension
can be found in \cite{\eho}.
The conformal dimensions of the last rational model of
type $\w(2,4,6)$  have been calculated in \cite{\beyond}.

As it will turn out in the next section the
first five rational models in Table~2 exhibit some interesting
analogy:  The representations of $\SL$ afforded by their
conformal characters belong to one and the same series $\rho_l$
(cf. \S 4.4 for details). So one could ask whether there exist
more rational models whose representations of $\SL$
belongs to the series $\rho_l$.
A more detailed investigation of the fusion algebras associated
to such potentially existing models showed that this is not
the case \cite{\ehf} (cf. also the speculation in \cite{\eho}).


\head
4. Uniqueness of conformal characters of certain rational models
\endhead

\subhead
4.1 Statement of the main theorem
\endsubhead
\proclaim{Main theorem}
Let $c$ be any of the central charges of Table 1 or 2, let $H_c$
denote the set of corresponding conformal dimensions, and let
$H$ be a subset of $H_c$ containing $0$. Assume that there exist
nonzero functions $\xi_{c,h}$ ($h\in H$), holomorphic on the upper half
plane, which satisfy the following conditions:
\roster
\item
The functions $\xi_{c,h}$ are modular functions for some
congruence subgroup of $\SL=\SLZ$.
\item
The
space of functions spanned by the $\xi_{c,h}$ ($h\in\ H$)
is invariant under $\SL$ with respect to the action
$(A,\xi)\mapsto\xi(A\tau)$.
\item
For each $h\in H$ one has $\xi_{c,h}=\LO(q^{-\tilde c/24})$
as $\Im(\tau)$  tends to infinity, where $\tilde c = c -24 \min H$.
\item
For each $h\in H$ the function $q^{-{(h-\frac c{24})}}\xi_{c,h}$
is periodic with period 1.
\item
The Fourier coefficients of the $\xi_{c,h}$ are rational numbers.
\endroster
Then $H=H_c$, and, for each $h\in H$, the function $\xi_{c,h}$
is unique up to multiplication by a scalar.
\endproclaim
\remark{Remarks}
1. Note that the theorem only ensures the uniqueness of the functions
$\xi_{c,h}$ but not their existence. However, they do indeed exist.
For Table 1 the existence of the corresponding functions is a well-known
fact \cite{\capelli, \eho}: explicit formulas for them can be given
in terms of the Riemann-Jacobi theta series
$$\sum\Sb x\in\Z\\x\equiv\lambda\bmod 2k\endSb\exp(2\pi i\tau x^2/4k).$$
The existence of the functions $\xi_{c,h}$ related to Table 2 will be
proved elsewhere \cite{\UNS}.

2. Note that the conformal characters $\chi_{M}$ of a rational model
with $H$ as set of conformal dimensions satisfy the properties listed
under (2) -- (5) by the very definition of rational models and Zhu's
theorem  if we set  $\xi_{c,h}=\chi_{M}$ ($h=$ conformal dimension of
$M$). Property (1) is not part of this definition, and it is not clear
whether it is implied by the axioms for rational models. However,
there is evidence that it always holds true (cf\. the discussion below).

3. If we assume for a rational model corresponding to a row in Table 1
or Table~2 that its conformal characters satisfy (1) we can conclude
from our theorem that the corresponding set $H_c$ is exactly
the set of its conformal dimensions and that the properly normalized
functions  $\xi_{c,h}$  ($h\in H_c$) are its conformal characters.

4. For the proof of the theorem for the first 5 models of Table 2 the
assumption $0\in H$ is not needed, and it can possibly be dropped in
all cases. However, we did not pursue this any further: From the
physical point of view the assumption $0\in H$ is natural since $h=0$
corresponds to the vacuum representation of the underlying $\w$-algebra,
i.e.\ the representation
given by the $\w$-algebra itself.
\endremark

For the first two cases of Table 2 the requirement that the $\xi_{c,h}$
are modular functions on some congruence subgroup is not necessary. Here
we have the
\proclaim{Supplement to the main theorem}
For $c=-\frac85$ and $c=\frac 45$ and with $H_c$ as in Table 2 the
equality $H=H_c$ and the uniqueness of the $\xi_{c,h}$ ($h\in H$)
are already implied by properties (2) to (5).
\endproclaim

For the other cases we do not know whether the statement about
the uniqueness of $H$ and the $\xi_{c,h}$ remains true if one takes
also into  account non-modular functions or non-congruence subgroups.

However, as already mentioned, it seems to be reasonable to expect that
the conformal characters associated to rational models satisfy (1).
Evidence for this is given by the following:

There is no example of a conformal character of any rational model
which is not a modular function on a  congruence subgroup.

As mentioned above the functions $\xi_{c,h}$, whose uniqueness is
ensured by the Main theorem, exist. As it turns out they can be
normalized so that their Fourier coefficients are always nonnegative
integers (for the case of Table 2 cf\. \cite{\UNS}). This gives
further evidence that they are identical with the conformal characters
of the corresponding $\w$-algebra models whence the latter
therefore satisfy (1).

According to the Main theorem, for each $H_c$ of Table 1 and 2
the $\SL$-module spanned by the $\xi_{c,h}$ is uniquely determined.
In particular the $S$-matrix (i.e\. the matrix representing the action
of $S$ with respect to the basis given by the $\xi_{c,h}$  with the
normalization indicated in the preceding remark) is unique. Closed
formulas for the $S$-matrices  corresponding to the first four rows
of Table 2 can be found in \cite{\UNS}. They can be compared with the
$S$-matrix of the corresponding $\w(2,4)$ rational model with
$c=-\frac{444}{11}$ as numerically computed in \cite{\ehd} using
so-called direct calculations in the $\w$-algebra. Both $S$-matrices
coincide within the range of the numerical precision.

All rational models listed in Table 2 are minimal models of Casimir
$\w$-algebras for which formulas for the corresponding conformal
characters have been obtained in \cite{\kac} under the assumption of a certain
conjecture.
Once more, the conformal characters so obtained
are modular functions on congruence subgroups \cite{\UNS}.

In the rest of \S 4 we prove our main theorem. To this end we will
develop some general tools dealing with modular representations, i.e\.
with representations of $\SL=\SLZ$ on spaces of modular functions or forms.
These methods are introduced in the
next two subsections. In \S 4.4 we conclude with the proof of the main theorem.

\subhead
4.2 A dimension formula for spaces of vector valued modular forms
\endsubhead

In this section we state dimension formulas for spaces of vector
valued modular forms on $\SLZ$.
These formulas are one of
the main tools in
the proof of the main theorem. It is quite natural in the context of conformal
 characters, or more generally in the context of modular representations, to
 ask for such formulas: The vector $\chi$ whose entries are the conformal
characters of a rational model, multiplied by a suitable power of $\eta$,
is exactly what we shall call a vector valued modular form, and is as such
an element of a finite dimensional space. (The latter holds true at least
in the case where the characters are invariant under a subgroup of finite
index in $\SL$; see the assumptions in the theorem below).

Multiplying $\chi$ by an odd power of $\eta$ yields a vector valued modular
form of half-integral weight. However, because of the ambiguity of the
squareroot of $c\tau+d$ ($c,d$ being the lower entries of a matrix in
$\SL$) we now do not deal with a vector valued modular form on $\SLZ$
but rather on a certain double cover $\CSL=\CSLZ$ of this group.

We now make these notions precise.

The double cover $\CSL$ is defined as follows: the group elements are
the pairs $(A,w)$, where $A$ is a matrix in $\SL$ and $w$ is a holomorphic
function on $\PH$ satisfying $w^2(\tau)=c\tau+d$ with $c,d$  the
lower row of $A$. The multiplication of two such pairs is defined by
$$(A,w(\tau))\cdot(A',w'(\tau))=(AA',w(A'\tau)\cdot w'(\tau)).$$

For any $k\in\Z$ we have an action of $\CSL$ on functions $f$ on $\PH$
given by
$$(f|_k(A,w))(\tau)=f(\tau)\,w(\tau)^{-2k}.$$
Note that for integral $k$ this action factors to an action of $\SL$,
which is nothing else than the usual ``$|_k$''-action
of $\SL$ given by $(f|_kA)(\tau)=f(\tau)(c\tau+d)^{-k}$.

For a subgroup $\Delta$ of $\SL$ we will denote by $D\Delta \subset \CSL$
the preimage of $\Delta$ with respect to the natural projection $\CSL@>>>\SL$
mapping elements to their first component.

Special subgroups of $D\Gamma$ which we have to consider below are the groups
$$ \Gamma(4m)^{\sharp} = \{ (A,j(A,\tau)) \vert A\in \Gamma(4m) \}.$$
Here, for $A\in\Gamma(4m)$, we use
$$ j(A,\tau) = \vartheta(A\tau)/\vartheta(\tau)$$
where $\vartheta(\tau) = \sum_{n\in\Z} q^{n^2}$.
It is well-known that indeed $j(A,\tau) = \epsilon(A) \sqrt{c\tau+d}$
where $c,d$ are the lower row of $A$ and $\epsilon(A) = \pm 1$.
Explicit formulas for $\epsilon(A)$ can be found in the literature, e.g\.
\cite{\skoruppa}.

We can now define the notion of a vector
valued modular form on $\SL$ or $\CSL$.
\definition{Definition}
For any representation $\rho\colon\CSL\rightarrow \GL{n}$ and any
number $k\in\frac12\Z$ denote by $M_k(\rho)$ the space of all holomorphic
maps $F\colon\PH@>>>\C^n$ which satisfy $F|_k\alpha=\rho(\alpha)F$ for all
$\alpha\in\CSL$, and which are bounded in any region $\Im(\tau)\ge r>0$.
Denote by $S_k(\rho)$ the subspace of all forms $F(\tau)$ in $M_k(\rho)$
which tend to 0 as $\Im(\tau)$ tends to infinity.
\enddefinition

If $\rho$ is a representation of $\SL$ and $k$ is integral we use $M_k(\rho)$
for $M_k(\rho\circ\pi)$, where $\pi$ is the projection of $\CSL$ onto the first
component. Clearly, in this case the transformation law for the functions $F$
of
$M_k(\rho)$ is equivalent to $F|_kA=\rho(A)F$ for all
$A\in\SL$. In general, if $k$ is integral, the group $\CSL$ may be replaced
by $\SL$ in all of the following considerations.

Finally, for a subgroup $\Delta$ of $\CSL$ or $\SL$ we use $M_k(\Delta)$ for
the space of modular forms of weight $k$ on $\Delta$ in the usual sense. In the
case
$\Delta\subset\SL$ the weight $k$ has of course to be integral.
The reader may not mix the two kinds of spaces $M_k(\rho)$ and $M_k(\Delta)$;
it will always be clear from the context whether $\rho$ and $\Delta$
refer to a representation or a group.

Clearly, if the image of $\rho$ is finite, i.e\. if the kernel
of $\rho$ is of finite index in $\CSL$ then the components of
an $F$ in $M_k(\rho)$ are modular forms of weight $k$ on this
 kernel. In particular, the space $M_k(\rho)$ is then finite
dimensional. Formulas for the dimension of these spaces
can be obtained as follows: Let $V$ be the complex vector space
 of row vectors of length $n=\dim\rho$, equipped with the $\CSL$-right
 action $(z,\alpha)\mapsto \rho(\alpha)^t z$, where ${()}^t$ means
transposition.
The space $M_k(\rho)$ can then be identified with the space
$\Hom_{\CSL}(V,M_k(\Delta))$ of $\CSL$-homomorphisms from $V$
to $M_k(\Delta)$, where $\Delta=\ker\rho$, via the correspondence
$$M_k(\rho)\ni F\mapsto\text{ the map which associates }z\in V
\text{ to }z^t\cdot F\in M_k(\Delta).$$
By orthogonality of group characters the dimension of
$\Hom_{\CSL}(V,M_k(\Delta))$ can be expressed in terms of the traces
 of the endomorphisms defined by the action of elements of $\CSL$
on $M_k(\Delta)$. These traces in turn can be explicitly computed
by using the Eichler-Selberg trace formula. In this way one can derive
the following theorem (cf\. \cite{\skoruppa, pp\. 100} for a complete proof):

\proclaim{Theorem (Dimension formula \cite{\skoruppa})}
Let  $\rho:\CSLZ\rightarrow\operatorname{GL}(n,\C)$ be a representation with
finite image and such that $\rho((\epsilon^2\id,\epsilon))=\epsilon^{-2k}\id$
for all fourth roots of unity $\epsilon$. and let $k\in\frac12\Z$. Then the
dimension
of $M_k(\rho)$ is given by the following formula
$$\align
\dim M_k(\rho)-\dim S_{2-k}(\overline\rho)=&\frac{k-1}{12}\cdot n
+\frac14\Re\big(\e^{\pi ik/2}\tr\rho((S,\sqrt{\tau}))\big)\\
&+\frac2{3\sqrt3}\Re\big(\e^{\pi i(2k+1)/6}\tr\rho((ST,\sqrt{\tau+1}))\big)\\
&+\frac12a(\rho)
-\sum_{j=1}^n\B_1(\lambda_j).
\endalign$$
Here the $\lambda_j$ ($1\le j\le n$) are complex numbers such that
$\e^{2\pi i\lambda_j}$ runs through the eigenvalues of  $\rho(T)$,
we use $a(\rho)$ for the number of $j$ such that $\e^{2\pi i\lambda_j}=1$,
 and we use $\B_1(x)=x'-1/2$ if $x\in x'+\Z$ with $0<x'<1$, and $\B_1(x)=0$
for $x$ integral.
Moreover, for $\tau\in\PH$, we use $\sqrt{\tau}$ and $\sqrt{\tau+1}$ for
those square roots which have
positive real parts.
\endproclaim
\remark{Remark}
For $k\ge 2$ the theorem gives an explicit formula for $\dim M_k(\rho)$
since in this case $\dim(S_{2-k}(\rho))=0$ (the components of a vector
valued modular form are ordinary modular forms on $\ker\rho$, and there
 exist no nonzero modular forms of negative weight and no cusp forms of weight
0).
\endremark

For $k=1/2$, $3/2$ and $\ker(\rho)\supset\Gamma(4m)^{\sharp}$ it is still
possible to give an explicit formula for $M_k(\rho)$ \cite{\skoruppa}.
However, we do not need those dimension formulas in full generality
but need only the following consequence of them:
\proclaim{Supplement to the dimension formula \cite{\skoruppa}}
Let $\rho:\CSLZ\rightarrow\operatorname{GL}(n,\C)$ be an
irreducible representation with $\Gamma(4m)^{\sharp} \subset \ker(\rho)$
for some integer $m$.  Then one has
$ \dim(M_{1/2}(\rho))= 0,1.$
Furthermore, if $\dim(M_{1/2}(\rho))=1$ then
the eigenvalues of $\rho(T)$ are of the form
$e^{2\pi i \frac{l^2}{4m}}$ with integers $l$.
\endproclaim
\remark{Remark}
A complete list of all those representations $\rho$ for which
$\dim(M_k(\rho))=1$ can be found in \cite{\skoruppa}.
\endremark

A proof of this supplement can be found in \cite{\skoruppa}.
It uses a theorem of Serre-Stark describing explicitly the modular forms
of weight $1/2$ on congruence subgroups.

\subhead
4.3  Three basic lemmas on representations of $\SLZ$
\endsubhead

In this section we will prove some lemmas which are useful for identifying
a given representation $\rho$ of $\SL$ if one has certain informations about
$\rho$, which can e.g\. easily computed from the central charge and the
conformal dimensions of a rational model.

Assume that the conformal characters of a rational model
are modular functions on some a priori unknown congruence subgroup.
Then the first step for determining the representation $\rho$, given by
the action of $\SL$ on the the conformal characters,
consists in finding a positive integer $N$ such that
$\rho$ factors through $\Gamma(N)$.
The next theorem tells us that the optimal choice of $N$ is given by the
order of $\rho(T)$.

\proclaim{Theorem (Factorization criterion)}
Let $\rho\colon\SL@>>>\GL{n}$ be a representation, and let $N>0$ be an
integer. Assume that $\rho(T^N)=1$, and, if $N>5$, that the kernel  of
$\rho$ is a congruence subgroup. Then $\rho$ factors through a
representation of $\SL/\Gamma(N)$.
\endproclaim
\demo{Proof}
The kernel $\Gamma'$ of $\rho$ contains the normal hull in $\SL$ of the
subgroup generated by $T^N$. Call this normal hull $\Delta(N)$. By a
result of \cite{\wohl} (but actually going back to Fricke-Klein) one has
$\Delta(N)=\Gamma(N)$ for $N\le 5$. If $N>5$ then by assumption we
have $\Gamma'\supset\Gamma(N')$ for some integer $N'$. Thus $\Gamma'$
contains $\Delta(N)\Gamma(NN')$, which, once more by \cite{\wohl},
 equals $\Gamma(N)$.
\enddemo

By the last theorem the determination of the representation $\rho$ associated
to a rational model with modular functions as conformal characters
is reduced to the investigation of the finite list of irreducible
representations of $\SL/\Gamma(N) \approx \MG{N}$ with some easily computable
$N$.
The following theorem, or rather its subsequent corollary, allows to reduce
this list dramatically.

\proclaim{Theorem ($K$-Rationality of modular representations)}
Let $k$ and $N>0$ be integers, let $K=\Q(\e^{2\pi i/N})$. Then the
$K$-vector space $M_k^K(\Gamma(N))$ of all modular forms on $\Gamma(N)$
of weight $k$ whose Fourier developments with respect to $\e^{2\pi i\tau/N}$
have coefficients in $K$ is invariant under the action $(f,A)\mapsto f|_k A$
of $\SL$.
\endproclaim
\demo{Proof}
Let $j(\tau)$ denote the usual $j$-function, which has Fourier coefficients
 in $\Z$ and satisfies $j(A\tau)=j(\tau)$ for all $A\in\SL$. Assume that $k$
is even. Then the map $f\mapsto f/{j'}^{k/2}$ defines an injection of the
$K$-vector space $M_k^K(\Gamma(N))$ into the field of all modular functions
on $\Gamma(N)$
whose Fourier expansions have coefficients in $K$. It clearly suffices to
show that the latter field is invariant under $\SL$. A proof for this can
be found in \cite{\Shi, p.\ 140, Prop\. 6.9 (1), equ\. (6.1.3)}.
The case $k$ odd can be reduced to the case $k$ even by
considering the squares of the modular forms in $M_k^K(\Gamma(N))$.
\enddemo

\proclaim{Corollary}
Let $\rho:\SL\rightarrow\operatorname{GL}(n,\C)$ be
a representation whose kernel contains $\Gamma(N)$ for some positive
integer $N$, and let $K=\Q(\e^{2\pi i/N})$. If, for some integer $k$,
there exists a nonzero element in $M_k(\rho)$ whose Fourier development
has Fourier coefficients in $K^n$, then $\rho(\SL)\subset\GLK{n}$.
\endproclaim
\remark
If one assumes that a vector valued modular form is related to the conformal
characters of a rational model which are modular functions of some congruence
subgroup then obviously all the Fourier coefficients are rational so that
the corollary applies.
\endremark
\demo{Proof}
If $F\in M_k(\rho)$ has Fourier coefficients in $K^n$ then $F|_kA$, by the
preceding theorem, has Fourier coefficients in $K^n$ too. Here $A$ is any
element in $\SL$. From $F|A=\rho(A)F$ we deduce that $\rho(A)$ has entries
in $K$.
\enddemo

\subhead
4.4 Proof of the main theorem
\endsubhead
We will now prove our main theorem stated in \S 4.1.
Pick one of the central charges $c$ in Table 1 or Table 2.
Assume that for some $H\subset H_c$ containing $0$ there exist functions
$\xi_{c,h}$ ($h\in H$) which satisfy the properties (1) to (5) of the main
theorem. Let $\xi$ denote the vector whose components are the functions
$\xi_{c,h}$ ordered with increasing $h$. Note that the $h$-values are pairwise
different modulo 1. By (3) the $\xi_{c,h}$ are thus linearly independent.
Hence,
we have a well-defined $|H|$-dimensional representation $\rho$ of the modular
group if we set $\xi( A\tau)=\rho(A)\xi(\tau)$ for $A\in\SL$. Finally, recall
that the Dedekind eta function $\eta$ is a modular form of weight $1/2$ for
$\CSL$,
 more precisely, that there exists a one-dimensional representation $\theta$ of
$\CSL$ on the group of 24-th roots of unity such that $\eta \in
M_{\frac12}(\theta)$.

For any half integer $k\in \frac12\Z$ such that
$$k\ge \tilde c/2$$
we have
$F:=\eta^{2k}\xi\in M_k(\rho\otimes\theta^{2k})$, as is
immediate from property (3) and the assumption that the $\xi_{c,h}$ are
holomorphic in the upper half plane. Let $k$ be the smallest possible half
integer
satisfying this inequality. The actual value is given in Table 3 below.

We shall show that by property (1) to (5) the representation $\rho$ is uniquely
determined (up to equivalence) and, in fact, is conjugate to the representation
in the last column of Table 3, respectively (notations will be explained
below).
In particular, $\rho$  has dimension equal to the cardinality of $H_c$, and
hence
we conclude $H=H_c$. The $h$-values are pairwise incongruent modulo 1, i.e.\
$\rho(T)$ has pairwise different eigenvalues. Since  $\rho(T)$ is a diagonal
matrix the representation $\rho$ is thus unique
up to conjugacy by diagonal matrices.

Finally, the kernel of $\rho$ is a congruence subgroup by property (1).
In particular, $\rho\otimes\theta^{2k}$ has a finite image. Thus we can
apply the dimension formulas stated in \S 4.2. It will turn out that
$M_k(\rho\otimes\theta^{2k})$ is one-dimensional.
Thus, if there actually exist functions $\xi_{c,h}$ satisfying (1) to (5) then
$M_k(\rho\otimes\theta^{2k})=\C\cdot \xi\eta^{2k}$. Since $\rho$ is unique up
to conjugacy by diagonal matrices we conclude that $\xi$ is unique up to
multiplication by such matrices, and this proves the theorem. We now give the
details.

\subhead
Determination of the representation $\rho$
\endsubhead
We first determine the equivalence class of the representation $\rho$.

Denote by $d$ the lowest common denominator of the rational numbers $h-c/24$
($h\in H_c$), and for any integer $k'$ let $l(k')$ be the lowest common
denominator of the numbers $h-c/24+k'/12$ ($h\in H_c$), i.e\. let
$$l(k')=12d/\gcd(12,k'd).$$
Clearly, the order
of $(\rho\otimes\theta^{2k'})(T)$ divides
$l(k')$. Let $k'$ the smallest nonnegative integer such that
$l=l(k')$ is minimal, and set $\rhot=\rho\otimes\theta^{2k'}$. The values of
$k'$ and $l$ are given in Table 3.

Note that $k'$ integral implies that $\rhot $ can be regarded as a
representation of $\SL$ (rather than $\CSLZ$). By property (1) its
kernel is a congruence subgroup. Thus we can apply the factorization
criterion of \S 4.3 to conclude that this kernel contains $\Gamma(l)$.
Note that here the assumption (1), namely that the $\xi_{c,h}$, are
invariant under a congruence subgroup, is crucial if $l>5$. For $l\le 5$,
this assumption is not necessary, which explains the supplement to the main
theorem.
\eject
\centerline{Table 3: Representations of $\SL$ and weights related
                     to certain rational models}
\smallskip\noindent
\centerline{
\vbox{\offinterlineskip
\def\tablespace{ height2pt&\omit&&\omit&&\omit&&\omit&&\omit&&\omit&\cr }
\def\tablerule{ \tablespace
                \noalign{\hrule}
                \tablespace      }
\hrule
\halign{&\vrule#&\strut\kern2pt\hfil$#$\hfil\kern2pt\cr
\tablespace
& \w\text{-algebra}&& c && k && k' && l &&\rhot=\rho\otimes\theta^{2k'} &\cr
\tablerule
&\w(2)&&1-6\frac{(p-q)^2}{pq}&&\frac12&&2&&8pq
  &&\sigma_p^q\otimes\sigma_q^p\otimes D_8^{pq}&\cr
\tablerule
&\w(2,\frac{(m-1)(q-2)}{2})&&1-3\frac{(2m-q)^2}{mq}
  &&\frac12&&\frac{1-3mq}2\bmod{12}&&mq
  &&\sigma_q^{2m}\otimes\tau_m^{2q}&\cr
\tablerule
&\w(2,q-3)&&1-\frac{(12-q)^2}{2q}&&\frac12&&-1-q\bmod 3&&16q
  &&\sigma_q^3\otimes D_{16}^q&\cr
\tablerule
&\w(2,q-5)&&1-\frac{(30-q)^2}{5q}&&\frac12&&\frac{1-5q}2 \bmod{12}&&5q
  &&\sigma_q^{30}\otimes\sigma_5^q&\cr
\tablerule
&\w_{G_2}(2,1^{14})&&-{8\over5}&&2&&4&&5&&\rho_5&\cr
\tablerule
&\w_{F_4}(2,1^{26})&&{4\over5}&&3&&10&&5&&\rho_5&\cr
\tablerule
&\w(2,4)&&-{444\over11}&&1&&6&&11&&\rho_{11}&\cr
\tablerule
&\w(2,6)&&-{1420\over17}&&1&&2&&17&&\rho_{17}&\cr
\tablerule
&\w(2,8)&& -{3164\over23}&&1&&10&&23&&\rho_{23}&\cr
\tablerule
&\w(2,4,6)&&-{13\over15}&&1&&1&&360
  &&\sigma_5^1\otimes D_8^0\otimes R_2(1,-)&\cr
\tablespace
}
\hrule}
}
Recall that in Table 3 the integers $p,q$ and $m$ are primes with $q\not=p,m$.
\hopp

We shall say that a representation of $\SL$ has level $N$ if its kernel
contains $\Gamma(N)$. Since any representation of level $N$ factors to a
representation of
$$\SL/\Gamma(N)\approx~\MG N,$$
it has a  unique decomposition as sum of irreducible level $N$
representations. Furthermore, there are only finitely many irreducible
level $N$ representation, and each such representation $\pi$ has a unique
product decomposition
$$\pi=\prod\Sb p^\lambda\Vert l\endSb\pi_{p^\lambda}$$
with irreducible level $p^\lambda$ representations  $\pi_{p^\lambda}$.
Finally, $\pi_{p^\lambda}(T)$ has order dividing $p^\lambda$, i.e\. its
eigenvalues are $p^\lambda$-th roots of unity. Since any $N$-th root of
unity $\zeta$ has a unique decomposition as product of the $p^\lambda$-th
roots of unity $\zeta^{\frac N{p^\lambda}x_p}$ with
$\frac N{p^\lambda}x_p\equiv 1\bmod p^\lambda$, we conclude:
\proclaim{Lemma}
Let $\zeta_j$ ($1\le j\le n=\dim\pi$) be the eigenvalues of $\pi(T)$. Then,
for each $p^\lambda\Vert N$, the eigenvalues $\not =1$ of $\pi_{p^\lambda}(T)$
(counting multiplicities) are exactly those among the numbers
$\zeta_j^{\frac N{p^\lambda}x_p}$ ($1\le j\le n$) which are not equal
to 1.
\endproclaim

\subhead
The representation $\rho$ in line 1 to 4 of Table 3
\endsubhead
First, we consider the rational models corresponding to the first 4 rows of
Table~3. By assumption $h=0$ is in $H$, i.e\. $\mu=\exp(2\pi i (-c/24+k'/12))$
is an eigenvalue of $\rhot (T)$. Let $\pi$ be that irreducible level $l$
representation in the sum decomposition of $\rhot $ such that $\pi(T)$ has
the eigenvalue $\mu$. Since $\pi$ is irreducible it has a decomposition
as product of irreducible representations $\pi_{p^\lambda}$ as above. Since
$\mu$
is a primitive $l$-th root of unity the lemma implies that the
$\pi_{p^\lambda}$
are nontrivial.

The minimal dimension of a nontrivial irreducible level $p^\lambda$
representation
is $2$, 3 or $(p-1)/2$ accordingly if $p^\lambda$ equals 8, 16 or is an odd
prime
\cite{\NobsW, p. 521ff}. Hence we have the inequalities
$$\dim\pi\ge
\cases
(p-1)(q-1)/2&\text{for row 1}\\
(m-1)(q-1)/4&\text{for row 2}\\
3(q-1)/2&\text{for row 3}\\
q-1&\text{for row 4}
\endcases.
$$
For row 1, 3 and 4 the right hand side equals the cardinality of $H_c$
respectively. In these cases we thus conclude that $\rhot =\pi$ is irreducible,
that it is equal to a product of  nontrivial level $p^\lambda$ representations
with
minimal dimensions, and, in particular, that $H=H_c$.

For row 2 the right hand side is smaller than the cardinality of $H_c$.
However, here we can sharpen the above inequality:
First we note that the level $p$ representations of dimension $(p-1)/2$
have parity $(-1)^{(p+1)/2}$, whence the product of the corresponding
level $m$ and $q$ representations has parity $(-1)^{(mq-1)/2}$. On the
other hand any irreducible subrepresentation has the same parity $\rhot $,
 i.e\. the parity $(-1)^{k'}=(-1)^{(mq+1)/2}$.  Hence $\pi$ cannot equal
a product of two nontrivial level $m$ and $q$ representations of minimal
dimension. The dimension of the second smallest nontrivial irreducible
level $p$ representations is $(p+1)/2$. Under each of these representations
$T$ affords eigenvalue 1. Since $T$ under $\rhot $ affords no $m$-th root of
unity as eigenvalue we conclude that $\pi$ cannot be equal to a product of a
$(q+1)/2$ dimensional level $q$ and a $(m-1)/2$ dimensional level $m$
representation. Thus, $$\dim \pi\ge (m+1)(q-1)/4.$$
The right hand side equals $|H_c|$, and we conclude as above that $H=H_c$,
that $\rho$ is irreducible, and that $\rhot $ equals a product of an
irreducible  $(q-1)/2$ dimensional level $q$ and an
irreducible $(m+1)/2$ dimensional level $m$ representation.

To identify $\rho$ it thus remains to examine the nontrivial level
$p^\lambda$ representations with small dimensions (cf\. \cite{\NobsW, p.
521ff}).

Let $p^\lambda=p$ be an odd prime. There exist exactly two irreducible
level $p$ representations with dimension $(p-1)/2$. The image of $T$
under these representations has exactly the eigenvalues
$\exp(2\pi i \varepsilon x^2/p)$ ($1\le x\le (p-1)/2$) where  for one of
them $\varepsilon$ is a quadratic residue modulo $p$, and a quadratic
non-residue for the other one \cite{\NobsW}. Call these representations
accordingly $\sigma_p^\varepsilon$. Similarly there exist exactly 2 irreducible
level $p$ representations with dimension $(p+1)/2$, denoted by
$\tau_p^{\varepsilon}$
(with $\varepsilon$ being a quadratic residue or non-residue modulo $p$). The
eigenvalues of $\tau_p^{\varepsilon}(T)$ are $\exp(2\pi i\varepsilon x^2/p)$
($0\le x\le (p-1)/2$).

Let  $p^\lambda=8$. There exist exactly 4 irreducible two dimensional level 8
representations which we denote by $D_8^x$ ($x$ being an integer modulo 4).
The eigenvalues of the image of $T$ under the representation $D_8^x$  are
$\exp(2\pi i(1+2x)/8)$ and $\exp(2\pi i(7+2x)/8)$.

Let $p^\lambda=16$. There are 16 irreducible three dimensional level 16
representations. These   can be distinguished by their eigenvalues
of the image of $T$. In particular, there are four of these representations,
denoted by $D_{16}^x$ ($x \bmod 4$), where the image of $T$ has the eigenvalues
$\exp(2\pi i (2x+3)/8),\exp(2\pi i (3x-6)/16),\exp(2\pi i(3x+2)/16)$.

Summarizing we find $\rhot =\sigma_p^{n_p}\otimes\sigma_q^{n_q}\otimes
D_8^{n_8}$,
$=\sigma_q^{n_q}\otimes\tau_m^{n_m}$, $=\sigma_q^{n_p}\otimes D_{16}^{n_{1}6}$
or
 $=\sigma_q^{n_q}\otimes\sigma_5^{n_5}$, respectively, with suitable numbers
$n_p,\dots$. The latter can be easily determined using the Lemma and the
description
of $H_c$ in Table 1. The resulting values are given in Table 3.

\subhead
The representation $\rho$ in line 5 to 9 of Table 3
\endsubhead
We now consider the rational models corresponding to row 5 to 9 of Table 3.
Here the level of $\rhot$ is a prime $l$, the dimension of $\rho$ is $\le l-1$,
and the eigenvalues of $\rho(T)$ are pairwise different primitive $l$-th root
of unity.

We show that $\rhot$ is irreducible with dimension $l-1$. Assume that
$\rhot$ is reducible or has dimension $<(l-1)$. The only irreducible
level $l$ representations with dimension $<(l-1)$ for which the image
of $T$ does not
afford eigenvalue 1 are $\sigma_l^\varepsilon$. Thus there only two
possibilities: (a) $\rhot=\sigma_l^\varepsilon$ or (b)
$\rhot=\sigma_l^\varepsilon\oplus\sigma_l^{\varepsilon'}$.
For $l=5,17$ the representations $\sigma_l^\varepsilon$ have parity
$-1$, whereas $\rhot$ has parity $+1$, a contradiction. For $l=11,23$
we note that $\xi\eta^2$ is an element of $M_1(\rhot\otimes\theta^{2-2k'})$.
We shall show in moment that the dimension of
$M_1(\sigma_l^\varepsilon\otimes\theta^{2-2k'})$ is 0, which gives the
desired contradiction (to recognize the contradiction in case (b) note
that the ``functor'' $\rho\mapsto M_k(\rho)$ respects direct sums).

Since the dimension formula gives explicit dimensions only for $k\not=1$ we
cannot apply it directly for calculating the dimension of
$M=M_1(\sigma_l^\varepsilon\otimes\theta^{2-2k'})$. For $l=11$ we note that
$\eta^2 M$ is a subspace of $M_2(\sigma_l^\varepsilon\otimes\theta^{4-2k'})$.
 To the latter we can apply the dimension formula, and find
(using $\tr \sigma_l^\varepsilon(S)=0$, $\tr \sigma_l^\varepsilon(ST)=-1$)
that its dimension is 0.
For $l=23$ and $\varepsilon=1$ we consider
$M_{3/2}(\sigma_l^1\otimes\theta^{3-2k'})$
which contains $\eta M$. We find that its dimension equals
$$\dim S_{1/2}(\sigma_l^{-1}\otimes\theta^{-(3-2k')})\le
 \dim M_{1/2}(\sigma_l^{-1}\otimes\theta^{-(3-2k')})\,,$$
which equals 0 by the
supplement in \S 4.2 (for applying the supplement note that
$\sigma_l^{-1}\otimes\theta^{-(3-2k')}$ has a kernel containing
$\Gamma(23\cdot 24)^\sharp$ and represents $T$ with eigenvalues
$\exp(2\pi i (-24x^2+17\cdot23)/23\cdot24)$ ).
Finally, by the dimension formula we find
$$\dim M_1(\sigma_l^{-1}\otimes\theta^{2-2k'})=
\dim S_1(\sigma_l^{1}\otimes\theta^{-(2-2k')}),
$$
and the right hand side equals 0 since
$\dim S_{3/2}(\sigma_l^{1}\otimes\theta^{-(1-2k')})=0$ by the supplement.

Thus, $\rhot$ is irreducible of dimension $l-1$, which implies in particular
$H=H_c$. There exist exactly $(l-1)/2$ irreducible level $l$ representations
of dimension $l-1$ \cite{\Dorn, p\. 228}. We now use property (5) of the main
theorem, which implies that the Fourier coefficients of $\xi\cdot\eta^{2k'}$
are rational. Hence, by the corollary in \S 4.3 we find that $\rhot$ takes
values in $\GLK{l-1}$ with $K$ being the field of $l$-th roots of unity.
There is exactly one irreducible level $l$ representations of dimension
$l-1$ whose character takes values in $K$ \cite{\Dorn, p\. 228}; denote it
by $\rho_l$. Then $\rhot=\rho_l$.

\subhead
The representation $\rho$ in line 10 of Table 3
\endsubhead
Finally, we consider the last rational model of Table 3. Here $\rhot$ has
level $360=8\cdot5\cdot9$. The eigenvalue of $\rhot(T)$ corresponding to $h=0$
is a primitive 360-th root of unity. Hence by the lemma there exists an
irreducible
 subrepresentation $\pi$ of $\rhot$ which factors as a product of nontrivial
irreducible representations of level 8, 5 and 9, respectively. The minimal
dimension of an irreducible nontrivial level 8, 5 or 9 representation is 2,
2 and 4, respectively \cite{\NobsW, p\. 521}. Thus $\dim\pi\ge 16=|H_c|$, and
hence $H=H_c$ and $\rhot=\pi$. The eigenvalues of $\rhot(T)$ can be read off
from Table 2. Using the lemma and the representations $D_8^x$ and
$\sigma_5^\varepsilon$ introduced above, we find
$$\rhot=D_8^{0}\otimes\sigma_5^{1}\otimes R$$
for an irreducible level 9 representation $R$ with dimension 4, which
represents $T$ with eigenvalues $\exp(2\pi i x^2)$ ($1\le x\le 4$), and is odd.
Looking up \cite{\NobsW} we find that there is exactly one such representation,
following \cite{\NobsW} we denote it by  $R_2(1,-)$.

\subhead
Computation of dimensions
\endsubhead
It remains to show $d=\dim M_k(\rhot\otimes\theta^{2k-2k'})\le1$.
For the first 4 rows of Table 3 this follows from the supplement in \S 4.2
and the irreducibility of $\rho$ (in fact it can be shown that $d=1$
\cite{\skoruppa}). For row 5 and 6 we find $d=1$ by the dimension formula and
using $\tr\rho_l(S)=0$, $\tr\rho_l(ST)=1$ (valid for arbitrary primes $l$). For
the remaining cases (where $k=1$) we multiply $M_1(\rhot\otimes\theta^{2-2k'})$
by $\eta$ for obtaining  $d'=\dim M_{3/2}(\rhot\otimes\theta^{3-2k'})$ as upper
bound. Again, using the dimension formula and its supplement we find $d'=1$.

This concludes the proof of the main theorem.\qed

\head
Acknowledgments
\endhead

W. E. would like to thank the working group of W. Nahm and D. Zagier for
many useful discussions.

All calculations have been done using the computer algebra package
PARI-GP~\cite{\gp}.

\hopp


\enddocument


\Refs
\refstyle{A}
\widestnumber\key{xxxxxxx}



\ref\key \bpz
\by A.A.\ Belavin, A.M.\ Polyakov, A.B.\ Zamolodchikov
\paper Infinite Conformal Symmetry in Two-Dimensional
       Quantum Field Theory
\jour Nucl. Phys. {\bf B}
\vol 241 \yr 1984 \pages 333-380
\endref

\ref\key \witten
\by E.\ Witten
\paper  Quantum Field Theory and the Jones Polynomial
\jour Commun. Math. Phys.
\vol 121 \yr 1989 \pages 351-399
\endref

\ref\key \cardy
\by J.L.\ Cardy
\paper Operator Content of Two-Dimensional
       Conformally Invariant Theories
\jour Nucl. Phys. {\bf B}
\vol 270 \yr 1986 \pages 186-204
\endref

\ref\key \bouwschou
\by P.\ Bouwknegt, K.\ Schoutens
\paper $\w$-Symmetry in Conformal Field Theory
\jour Phys. Rep.
\vol 223 \yr 1993 \pages 183-276
\endref

\ref\key \nrt
\by
W.\ Nahm, A.\ Recknagel, M.\ Terhoeven
\paper
Dilogarithm Identities in Conformal Field Theory
\jour  Mod. Phys. Lett. {A}
\vol 8 \yr 1993 \pages 1835-1847
\endref

\ref\key \roesgen
\by J.\ Kellendonk, M.\ R\"osgen, R.\ Varnhagen
\paper Path Spaces and $\w$-Fusion in Minimal Models
\jour   Int. Jour. Mod. Phys. {\bf A 9} (1994) 1009-1023
\endref

\ref\key \UNS
\by W.\ Eholzer, N.-P.\ Skoruppa
\paper  Conformal Characters and Theta Series
\jour in preparation
\endref


\ref\key \frenkel
\by I.B.\ Frenkel, Y.\ Huang, J.\ Lepowsky
\book  On Axiomatic Approaches to Vertex Operator Algebras and Modules
\bookinfo Memoirs of the American Mathematical Society, Volume 104, Number 494
\publ American Mathematical Society
\publaddr Providence, Rhode Island
\yr 1993
\endref

\ref \key \duke
\by I.B.\ Frenkel, Y. Zhu
\paper Vertex Operator Algebras Associated to Representations
       of Affine and Virasoro Algebras
\jour Duke Math. J.
\vol 66(1) \yr 1992  \pages 123-168
\endref

\ref\key \wer
\by W.\ Nahm
\book Chiral Algebras of Two-Dimensional Chiral
       Field Theories and Their Normal Ordered Products
\bookinfo proceedings of the Trieste Conference on
          ``Recent Developments in Conformal Field Theories''
\publ World Scientific
\publaddr Singapore
\yr 1989
\endref

\ref\key \feher
\by L.\ F\'eher, L.\ O'Raifeartaigh, I.\ Tsutsui
\paper The Vacuum Preserving Lie Algebra of a Classical $\w$-algebra
\jour Phys. Lett. {\bf B}
\vol 316 \yr 1993 \pages 275-281
\endref

\ref\key \zhu
\by Y.\ Zhu
\paper  Vertex Operator Algebras, Elliptic Functions, and Modular Forms
\jour Ph.D. thesis, Yale University, 1990
\endref

\ref \key \wang
\by W.\ Wang
\paper  Rationality of Virasoro Vertex Operator Algebras
\jour Int. Research Notices (in Duke Math. J.)
\vol 7 \yr 1993   \pages  197-211
\endref

\ref\key \eho
\by W.\ Eholzer, M.\ Flohr, A.\ Honecker, R.\ H{\"u}bel, W.\ Nahm,
    R.\ Varnhagen
\paper Representations of $\w$-Algebras with Two Generators
       and New Rational Models
\jour Nucl. Phys. {\bf B}
\vol 383 \yr 1992 \pages 249-288
\endref

\ref \key \andmoore
\by G.\ Anderson, G.\ Moore
\paper  Rationality in Conformal Field Theory
\jour Commun. Math. Phys.
\vol 117  \yr 1988   \pages  441-450
\endref


\ref\key \roc
\by A.\ Rocha-Caridi
\book Vacuum Vector Representations of the Virasoro Algebra
\bookinfo in 'Vertex Operatos in Mathematics and Physics',
          S. Mandelstam and I.M. Singer
\publ MSRI Publications Nr. 3, Springer
\publaddr Heidelberg
\yr 1984
\endref


\ref\key \capelli
\by A.\ Cappelli, C.\ Itzykson, J.B.\ Zuber
\paper The A-D-E Classification of Minimal and $A_1^{(1)}$
       Conformal Invariant Theories
\jour  Commun. Math. Phys.
\vol 113 \yr 1987 \pages 1-26
\endref

\ref\key \kac
\by E.\ Frenkel, V.\ Kac, M.\ Wakimoto
\paper Characters and Fusion Rules for $\w$-Algebras
       via Quantized Drinfeld-Sokolov Reduction
\jour Commun. Math. Phys.
\vol 147 \yr 1992  \pages 295-328
\endref


\ref\key \kacB
\by V.\ Kac
\book Infinite Dimensional Lie Algebras and Groups
\publ World Sientific
\publaddr Singapore
\yr 1989
\endref

\ref\key \beyond
\by W.\ Eholzer, A.\ Honecker, R.\ H{\"u}bel
\paper How Complete is the Classification of $\w$-Sym\-metries ?
\jour Phys. Lett. {\bf B}
\vol 308 \yr 1993 \pages 42-50
\endref

\ref\key \ehf
\by W.\ Eholzer
\paper Fusion Algebras Induced by Representations of the
       Modular Group
\jour  Int. Jour. Mod. Phys. {\bf A}
\vol 8 \yr 1993 \pages 3495-3507
\endref



\ref\key \ehd
\by W.\ Eholzer
\paper Exzeptionelle und Supersymmetrische $\w$-Algebren
       in Konformer Quantenfeldtheorie
\jour Diplomarbeit BONN-IR-92-10
\endref

\ref\key \skoruppa
\by N.-P.\ Skoruppa
\paper \"Uber den Zusammenhang zwischen Jacobiformen
       und Modulformen halbganzen Gewichts
\jour Bonner Mathematische Schriften
\vol 159 \yr 1985
\endref

\ref\key \wohl
\by K.\ Wohlfahrt
\paper An extension of F.\ Klein's level concept
\jour Illinois J. Math.
\vol 8 \yr 1964 \pages 529-535
\endref

\ref\key \Shi
\by G.\ Shimura
\book Introduction to the Arithmetic Theory of Automorphic Functions
\publ $\qquad$ Iwanami Sholten, Princeton Press
\publaddr Tokyo, Japan
\yr 1971
\endref


\ref \key \NobsW
\by A.\ Nobs
\paper Die irreduziblen Darstellungen der Gruppen $SL_2(Z_p)$
       insbesondere $SL_2(Z_2)$  I
\jour Comment. Math. Helvetici
\vol 51 \yr 1976 \pages  465-489
\endref

\ref \key \NobsW
\by A.\ Nobs, J.\ Wolfart
\paper Die irreduziblen Darstellungen der Gruppen $SL_2(Z_p)$
       insbesondere $SL_2(Z_2)$  II
\jour Comment. Math. Helvetici
\vol 51 \yr 1976 \pages  491-526
\endref

\ref\key \Dorn
\by L.\ Dornhoff
\book Group Representation Theory
\publ Marcel Dekker Inc.
\publaddr New York
\yr 1971
\endref



\ref \key \gp
\by C.\ Batut, D.\ Bernardi, H.\ Cohen, M.\ Olivier
\paper PARI-GP
\publ Universit\'e Bordeaux 1
\publaddr Bordeaux
\yr 1989
\endref
\endRefs
\enddocument